\documentclass[aps,prx,twocolumn,superscriptaddress]{revtex4-2}
\usepackage{amssymb,amsmath}
\usepackage{graphicx}
\usepackage{color}
\usepackage{hyperref}
\usepackage{listings}
\usepackage{siunitx}
\usepackage{gensymb}
\usepackage{textcomp}
\usepackage{xspace}
\usepackage[utf8x]{inputenc}

\hypersetup{
    unicode=false,          % non-Latin characters in Acrobat’s bookmarks
    pdftoolbar=true,        % show Acrobat’s toolbar?
    pdfmenubar=true,        % show Acrobat’s menu?
    pdffitwindow=false,     % window fit to page when opened
    pdfstartview={FitH},    % fits the width of the page to the window
    pdfauthor={none},     % author
    colorlinks=true,       % false: boxed links; true: colored links
    linkcolor=blue,          % color of internal links
    citecolor=red,        % color of links to bibliography
}

\begin{document}

%\preprint{}

\title{Trapping electrons in a room-temperature microwave Paul trap}


\author{Clemens Matthiesen}
\affiliation{Department of Physics, University of California, Berkeley, California 94720, USA}
\author{Qian Yu}
\affiliation{Department of Physics, University of California, Berkeley, California 94720, USA}
\author{Jinen Guo}
\affiliation{Department of Physics, University of California, Berkeley, California 94720, USA}
\author{Alberto M. Alonso}
% \affiliation{Department of Physics, University of California, Berkeley, California 94720, USA}
\author{Hartmut H\"{a}ffner}
\email[]{hhaeffner@berkeley.edu}
%\thanks{}
\affiliation{Department of Physics, University of California, Berkeley, California 94720, USA}

\begin{abstract}
We demonstrate trapping of electrons in a millimeter-sized quadrupole Paul trap driven at 1.6~GHz in a room-temperature ultra-high vacuum setup. Cold electrons are introduced into the trap by ionization of atomic calcium via Rydberg states and stay confined by microwave and static electric fields for several tens of milliseconds. A fraction of these electrons remain trapped longer and show no measurable loss for measurement times up to a second. Electronic excitation of the motion reveals secular frequencies which can be tuned over a range of several tens to hundreds of MHz. Operating a similar electron Paul trap in a cryogenic environment may provide a platform for all-electric quantum computing with trapped electron spin qubits.
\end{abstract}

%\keywords{}

\maketitle

\section{\label{Introduction}Introduction}
The spin up and down states of an electron form the archetypal two-level system in quantum physics and make the electron a natural candidate for realizing a quantum bit. Quantum computing approaches use electrons in both condensed matter and atomic systems, for instance confined in quantum dots or bound to donors in semiconductors~\cite{Watson2018, Yoneda2018a, Laucht2017}, or bound as valence electrons in trapped atomic ions~\cite{Brown2016, Bruzewicz2019}. In these examples, the confinement to either the host solid-state environment or to a much heavier ion can limit the potential of the electron spin qubit: for trapped ions, entanglement is typically mediated by the slow motion of the heavy ions in a shared trapping potential~\cite{Sorensen1999, Leibfried2003a}, which limits the gate speed, while in condensed matter systems unwanted coupling of the electron's charge and magnetic moment to the imperfect environment limits coherence times.

An approach which promises to remove these limitations is to confine individual free electrons in actual vacuum~\cite{Daniilidis2013electron,Peng2017,Kotler2017}. Here we show experimentally that this can be achieved with the type of traps used for the currently most advanced ion trap quantum computers, namely quadrupole Paul traps.
Compared to commonly trapped ions, the electron's charge-to-mass ratio is larger by a factor $10^4-10^5$, such that motion-based gates and shuttling operations could be sped up by two orders of magnitude.
Based on measurements for ion ground state qubits, which should experience similar decoherence mechanisms to trapped electron spin qubits, coherence times of at least a second are expected~\cite{Ruster2016}.
Furthermore, reducing the complex level structure down to the minimum of two levels rules out qubit errors due to population leakage~\cite{Brown2018}. Adapting the quantum-CCD architecture developed for trapped ions~\cite{Kielpinski2002CCD} to trapped electrons offers the opportunity to build a fast, modular, and high-fidelity quantum computer using advanced microwave technology~\cite{Mintert2001a, Piltz2016, Ospelkaus2011, Harty2016}, which promises better compatibility with current microfabrication methods compared to laser technology and optical beam delivery.

Beyond quantum computing, the experimental platform we introduce here may offer new avenues for creating and studying small cold plasma~\cite{Twedt2012}, highly controllable few- to single electron sources for electron optics applications~\cite{Kruit2016}, or single-electron mechanical oscillators~\cite{Burd2019}.

Trapping single electrons in vacuum has previously been achieved in two other platforms. First, electrons have been confined in cryogenic Penning traps in the early 1970s~\cite{Wineland1973} by combining a large magnetic field and a constant electric quadrupole field.
While several proposals have considered using single electrons in Penning traps as qubits~\cite{Ciaramicoli2003, Marzoli2009, Goldman2010}, limited work has been performed on experimental realisations so far~\cite{Bushev2008}.
Electrons can also be trapped above the surface of liquid helium, offering quantum information applications in milli-Kelvin environments~\cite{Lyon2006, Schuster2010} and recent experimental efforts have reached the single-electron regime~\cite{Koolstra2019a}.

Our approach to trapping electrons builds on the established quadrupole radiofrequency ion trap architecture, which is at the forefront of current quantum computing approaches with atomic ions~\cite{Brown2016, Bruzewicz2019}. Guiding electrons along a radiofrequency guide~\cite{Hoffrogge2011} has been achieved and electrons have been co-trapped with ions in a combined Paul and Penning trap~\cite{Walz1995}, but trapping electrons in a pure Paul trap has not been reported so far. While potential applications to quantum computing will require cryogenic environments~\cite{Daniilidis2013electron,Peng2017,Kotler2017}, we concentrate here on demonstrating electron trapping in a proof-of-principle experiment at room temperature.

\begin{figure}
\includegraphics[width = 3.375 in]{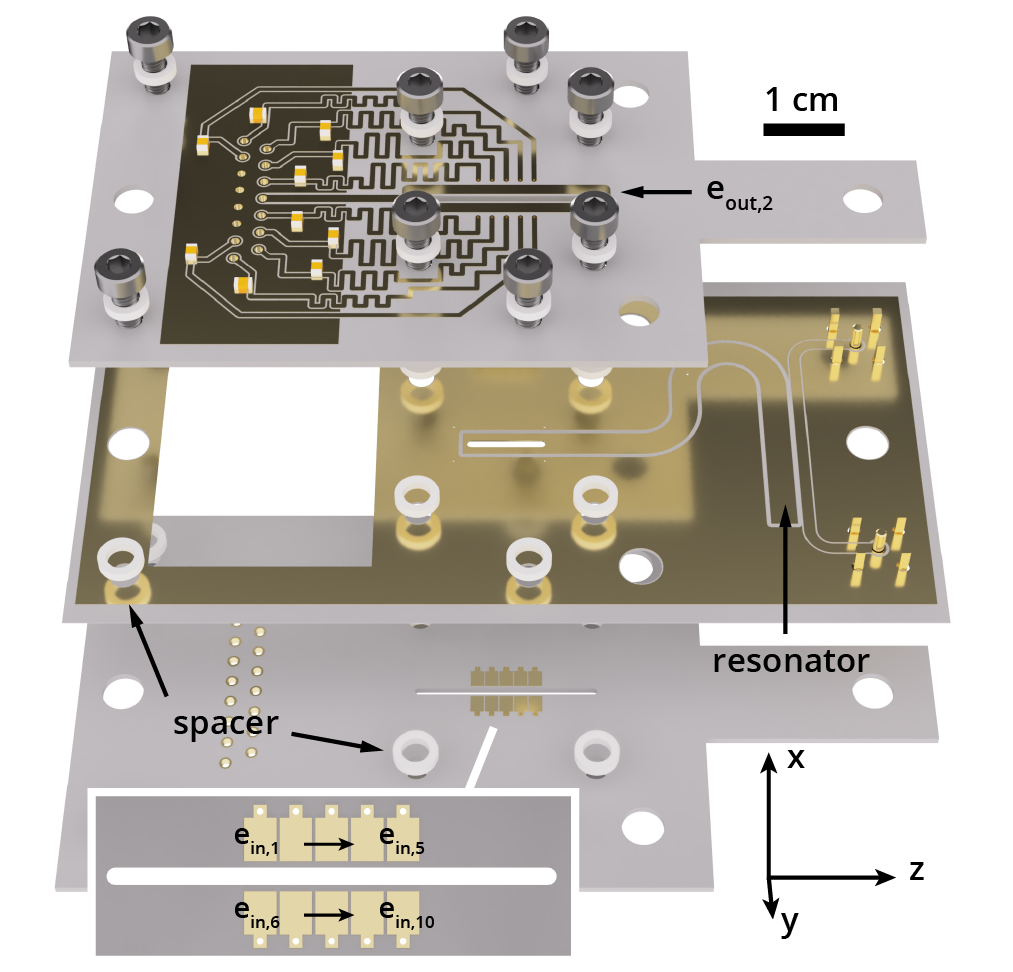}%
\caption{\label{fig:Fig1} Electron trap design. Exploded view of trap printed circuit boards.
Driving the halfwave co-planar waveguide resonator (central board)
gives rise to a quadrupole microwave trap inside the slot at the central end of the resonator. The two outside boards are identical and feature electrodes $e_{\mathrm{in,\;1-10}}$ ($e_{\mathrm{in,\;11-20}}$) on the bottom (top) to provide static confinement along the slot direction (see bottom inset for electrode labels). The boards are separated with alumina spacers of 1.27~mm height and have a footprint of about 5 by 10 cm.}
\end{figure}

Paul traps employ a rapidly oscillating quadrupole electric field to confine charged particles at the null of the quadrupole field in two or three dimensions. The effective confining potential can be described by the pseudopotential $U_{\mathrm{p}} = \frac{q^2 E^2}{4 m \Omega^2}$, where $E$ and $\Omega$ are the amplitude and frequency of the oscillating  electric field, and $q$ and $m$ the charge and mass of the trapped particle(s). The spatial dependence $U_{\mathrm{p}}(x, y, z)$ derives from the quadrupole electric field amplitude $E(x, y, z)$. The stability of trajectories for a charged particle in a quadrupole trap can be described with the $a_{\mathrm{M}}$ and $q_{\mathrm{M}}$ parameters of the Mathieu equation, which are known as the stability parameters in the context of ion traps, and have been studied theoretically and experimentally~\cite{Alheit1997,Leibfried2003review}. Typically, trajectories in the trap are stable if the frequency of motion $\omega$ of the particles in the potential is much slower than the frequency~$\Omega$ of the confining field, and if the pseudopotential depth, defined as the maximum of the pseudopotential, is much larger than the kinetic energy of the charged particles. While the Mathieu equation provides useful intuition for quadrupole traps, it should be noted that the pseudopotential picture and the treatment with the Mathieu equation are no longer accurate when the potential deviates from a purely harmonic dependence.

There are three main challenges to moving from ion to electron trapping. First, due to the lower electron mass the trapping field must be at higher frequencies. The stability parameters ($a_{\mathrm{M}}$ and $q_{\mathrm{M}}$) and depth of a quadrupole trap scale as $(m\Omega^2)^{-1}$, requiring the drive frequency~$\Omega$ for an electron trap to be about two orders of magnitude higher than for typical ion traps. Second, electrons must be created with energies low enough to stay confined by the trapping potential. We require both that cold electrons are injected directly into the trap center, and that the trap is sufficiently deep. Third, in the absence of fluorescence detection, we need a different mechanism to evidence trapping.

\section{\label{Experiment}Experiment design}
\subsection{\label{Trap}Electron trap}
We begin by describing the microwave quadrupole trap engineered for this experiment, shown in an exploded view in Fig.~\ref{fig:Fig1}. It consists of three double-sided printed circuit boards (PCBs) separated by alumina spacers. The central board features a co-planar $\lambda/2$ waveguide resonator capacitively coupled to a microwave feedline (right-hand side of the PCB). The end of the resonator in the board center contains a slot and functions as the trap's microwave electrode, providing an AC quadrupole field which confines electrons inside the slot in the $x$ and $y$-directions. The quality factor of the resonator is about 35. When fully assembled and connected inside the ultra-high vacuum (UHV) chamber, we measure a resonance frequency of $2\pi\times\;1.60$~GHz, and find we can reach about 100~V on the microwave resonator with 5~W input power. Integrating a co-planar resonator into the trap design provides a convenient solution to reaching the high frequencies needed for electron trapping and future cryogenic experiments can take advantage of previous work on waveguide resonators in the context of superconducting qubits~\cite{Goeppl2008, Malissa2013}. The resonator is held at DC ground potential via a tap in its center which connects it to the grounded top surface of the board. The outside PCBs, mirroring each other about the central board, each feature ten rectangular electrodes along the slot on the inside board surface, labeled $e_{\mathrm{in,\;j}}$ with $j=1-20$. Electrodes $e_{\mathrm{in,\;1-10}}$ are visible on the lower board and magnified in the inset, while electrodes $e_{\mathrm{in,\;11-20}}$ are on the hidden side of the upper board. The traces delivering voltages to the electrodes are on the outside surfaces, visible for the top board, and linked to a ground electrode via 10~pF decoupling capacitors. Both boards also feature a single electrode which surrounds the slot on the outside surface, and is labeled $e_{\mathrm{out,\;1}}$ for the bottom board and $e_{\mathrm{out,\;2}}$ for the top board. Electrodes $e_{\mathrm{in,\;1-20}}$ are used to apply a static quadrupole field, confining electrons in the $z$ (axial) direction, while $e_{\mathrm{out,\;1-2}}$ are held at DC ground potential. Wires soldered to the outer boards supply DC voltages in the $\pm 28$~V range from a 16 bit digital-to-analog-converter, while the microwave voltage is applied via SMA connectors to the central board.

\begin{figure*}
\includegraphics{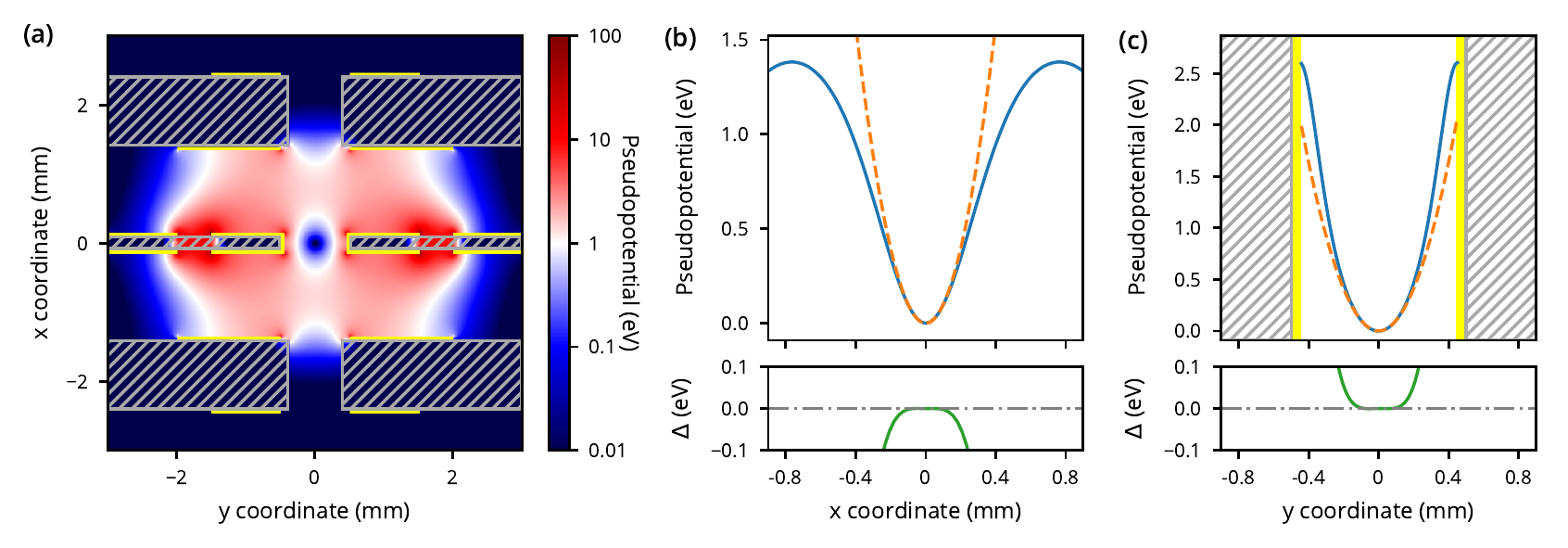}%
\caption{\label{fig:Fig2} Electron trapping potential. (a) Contour plot of the trapping potential based on the pseudopotential approximation in the $xy$-plane in the center of the slot for 90~V amplitude at 1.6~GHz frequency on the resonator electrode. The circuit board substrate is indicated by the hashed areas, metal electrodes are highlighted in yellow. (b) Top: pseudopotential along the $x$-axis through the trap center (blue continuous curve), compared to an ideal harmonic potential (dashed orange). Bottom: deviation $\Delta$ of pseudopotential from harmonic potential. (c) Top: pseudopotential along the $y$-axis through the trap center (blue continuous curve), compared to an ideal harmonic potential (dashed orange). Location of trap substrate (electrodes) shown as hashed grey (solid yellow) area. Bottom: deviation $\Delta$ of pseudopotential from harmonic potential.}
\end{figure*}

\begin{figure*}
\includegraphics{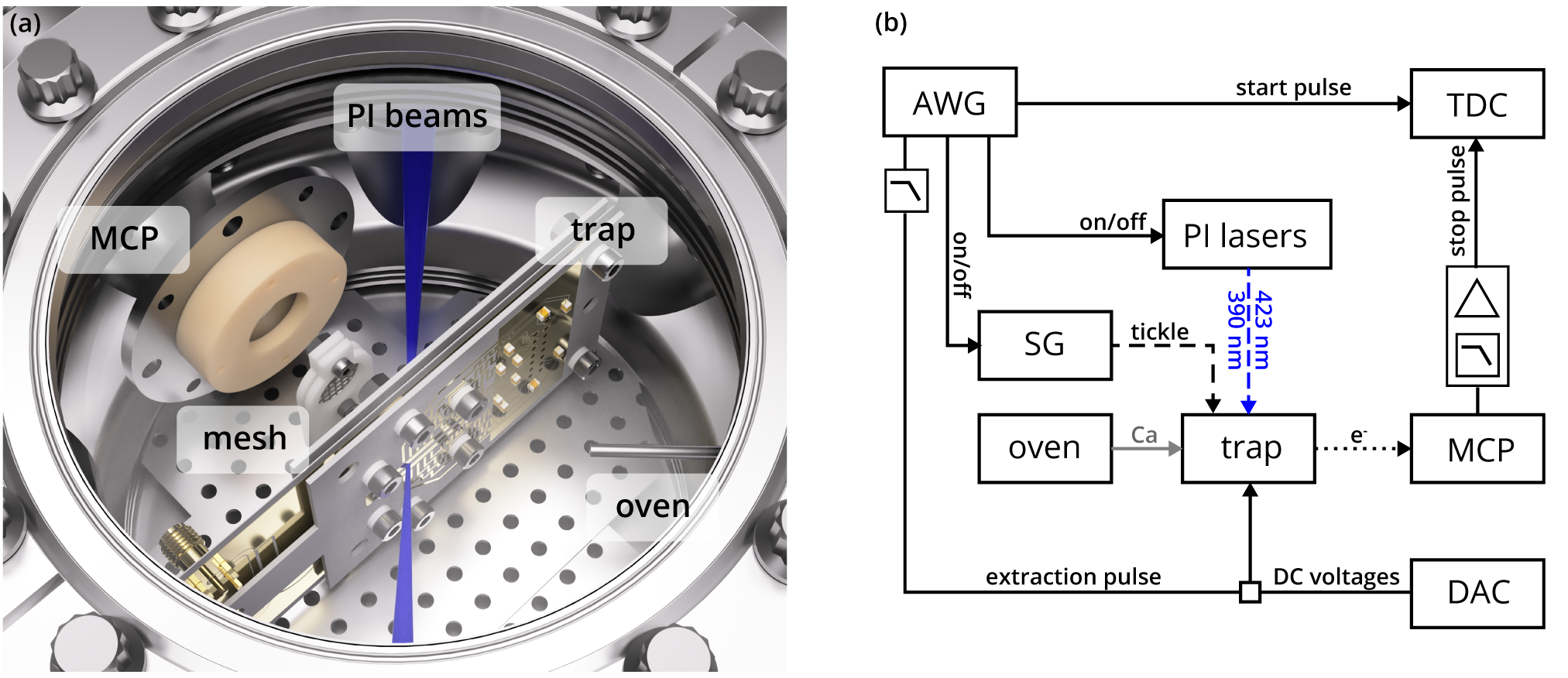}%
\caption{\label{fig:Fig3} Experimental setup and control schematic. (a) Simplified rendering of the setup inside the UHV chamber, showing the electron trap, a microchannel plate (MCP) detector, a mesh, the path of the PI beams, and an oven supplying calcium for ionization. (b) Key elements for experiment synchronization. An arbitrary waveform generator (AWG) provides the extraction pulses to the trap, and the start signal for a time-to-digital converter (TDC). It controls the timing for pulsing on and off the PI loading beams and a signal generator (SG), which excites the electron motion (`tickle'). Pulses from the MCP provide the stop Signals to the TDC. Low-pass filters prevent electronic pickup of the microwave trap drive and the extraction pulses by the MCP detection circuit which contains pulse shaping and amplification elements. The Ca oven and a digital-to-analog converter (DAC) for the DC trap voltages are operated with constant settings during an experiment.}
\end{figure*}

Fig.~\ref{fig:Fig2} details the trap pseudopotential experienced by an electron when the resonator supplies 90~V at $2\pi\times\;1.6$~GHz and all DC electrodes are grounded. Fig.~\ref{fig:Fig2}(a) displays a map of the pseudopotential for a cut through the trap in the $xy$-plane at the center of the slot, highlighting the trap substrate and electrodes as hashed grey and yellow areas, respectively.
In Fig.~\ref{fig:Fig2}(b) and (c) we show that the trap depth based on the pseudopotential approximation (continuous blue curves) is about 1.3~eV (or 15,000~K), limited by the weaker confinement along the $x$-direction. We compare the pseudopotential to an ideal harmonic potential (orange dashed curve) and find they match closely to a distance of about 100~$\mathrm{\mu m}$ from the trap center, as exemplified by the green curves in the bottom panels which show their difference $\Delta$. The secular frequency for an electron moving in this potential (radial modes of motion) corresponds to about $2\pi\times\;300$~MHz.

\subsection{\label{th_load}Electron loading and detection}
With a suitable trap design in place, we address the challenges of injecting electrons into the trap and detecting them. Previous experiments involving trapping electrons in Penning traps, or guiding electrons in a linear quadrupole potential employed electron guns, either as primary electron source~\cite{Hoffrogge2011}, or to create secondary electrons through collision ionization of background gas~\cite{Wineland1975,Walz1995}.
Here, we borrow the two-stage procedure for photoionization (PI) of calcium which is used for trapping ions from an atomic beam~\cite{Gulde2001}. It enables both the creation of very cold electrons by tuning the lasers close to the ionization threshold, and preferential ionization in the trapping region by optical alignment. Since fewer charged particles are introduced around the trap using this method, we also reduce accidental charging of the trap which would modify the trapping potential.
Detection is accomplished by applying voltage pulses to several DC electrodes which distort the trapping potential to extract trapped electrons, and accelerate them into a microchannel plate detector (MCP).

\subsection{\label{setup}Experiment setup and protocol}
The main components of the experimental setup and their alignment in the UHV chamber are shown in Fig.~\ref{fig:Fig3}(a), omitting electric leads for simplicity. The base pressure in the chamber is below $1\times10^{-10}$~mbar. In addition to the trap itself, the chamber contains a resistive oven aligned to direct an atomic calcium beam through the trap slots when heated (steel tube labelled `oven'), the two-stage MCP, and a steel mesh which directs electrons extracted from the trap towards the MCP. The PI laser beams (423 and 390~nm wavelength) are overlapped and traverse the chamber at right angles to the Ca oven, focusing near the trapping region with a beam waist of about 30~$\mathrm{\mu m}$. The 423~nm single-mode laser is tuned to be on resonance with the neutral calcium $4 ^{1}S_{0}$-$4 ^{1}P_{1}$ transition, while the free-running multi-mode 390~nm laser diode is tuned by temperature and current to maximise the electron ionization rate. While the $4 ^{1}P_{1}$-continuum ionization threshold is at about 389.8~nm, we find the ionization rate to peak when the diode center wavelength is about $390.3\pm0.2$~nm, suggesting ionization is taking place via Rydberg states~\cite{Gulde2001}. As such, we expect electrons to inherit only minimal kinetic energy from the ionization process and their energy in the trapping potential is rather determined by their ionization location and the phase of the microwave field. We note that since the PI lasers are co-propagating, Ca ionization conditions are met over an extended volume, which follows the laser beam path, and the majority of electrons are created outside of the trap. Constraints on optical access to the trap in our particular setup prevent us from aligning the two PI beams such that they only intersect inside the trap.

Fig.~\ref{fig:Fig3}(b) shows a schematic of the electronics setup for synchronizing the experiment. An arbitrary waveform generator (AWG) functions as the experiment clock, providing the trap extraction pulse, and the start signal triggering a time-to-digital-converter (TDC). The extraction pulses are added to the static voltages for DC confinement which originate from a digital-to-analog converter (DAC). The AWG further controls the timing for switching on and off both the 390-nm PI laser for loading and a signal generator (SG) used to apply an rf tone (labeled `tickle') to the trap. The electron detection signal is picked off from the MCP anode supply voltage with a high-pass filter, shaped and amplified so that it can be used as the TDC stop signal. Low-pass filtering the extraction pulses ($2 \pi \times 50$~MHz cut-off) prevents electronic pickup at the MCP and a further low-pass filter ($2 \pi \times 200$~MHz cut-off) in front of the shaping and amplification circuit removes pickup at the frequency of the microwave drive.

\begin{figure}
\includegraphics{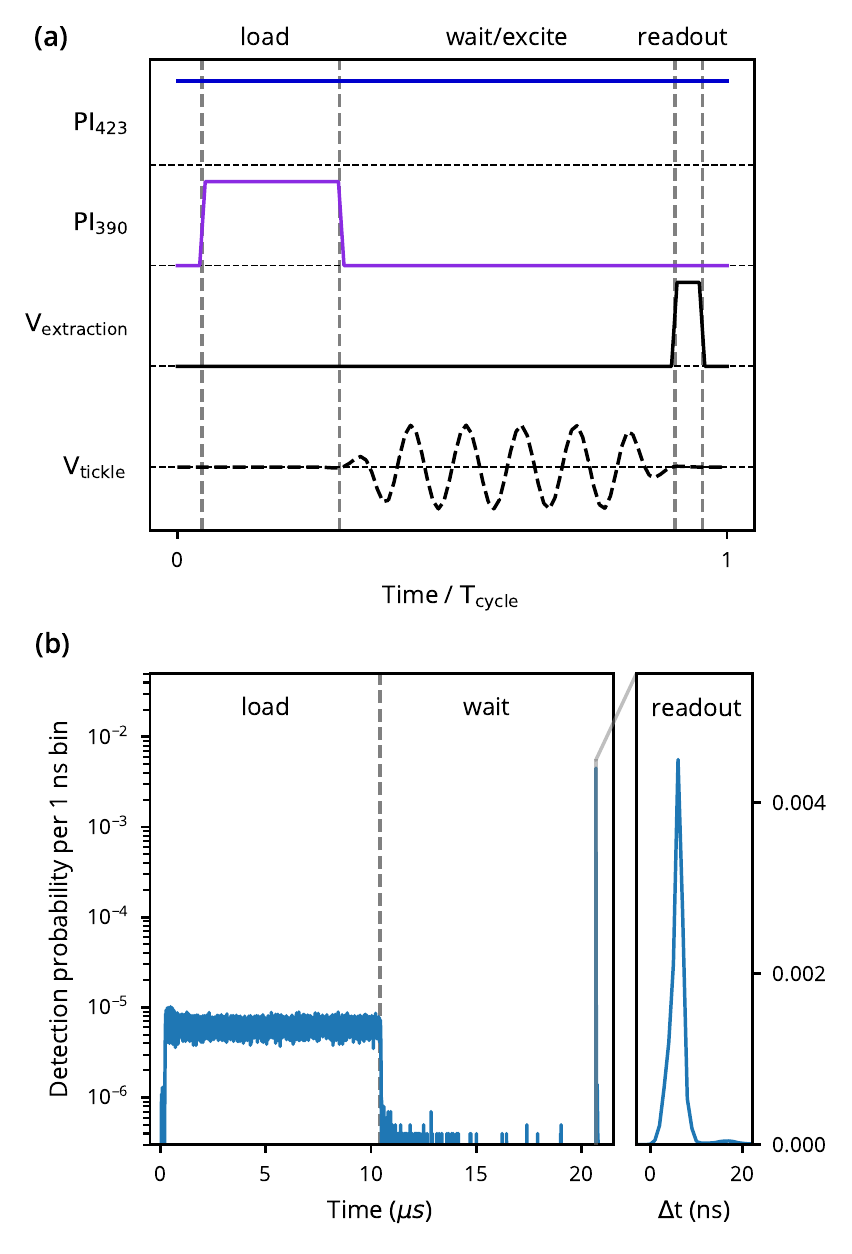}%
\caption{\label{fig:Fig4} Measurement protocol and typical data. (a) Illustration of one cycle of the experimental protocol. Free electrons are created during the loading phase when the 390-nm laser is switched on. An rf tone is applied to one DC electrode in some experiments during the `wait/excite' phase. Voltage pulses to three DC electrodes in the `readout' phase eject electrons in the direction of the mesh and MCP. (b) Histogram of MCP detection events for experiment with load and wait times $t_{\mathrm{load}}=t_{\mathrm{wait}}\approx 10~ \mathrm{\mu s}$. During loading, some untrapped electrons are accelerated into the MCP, replicating the 390~nm laser pulse shape. The extraction pulse empties the trap into the MCP, resulting in a large and sharply localized signal. Inset: Close-up of the histogram during the readout phase.}
\end{figure}

The timing for one experimental cycle is illustrated in Fig.~\ref{fig:Fig4}(a). It starts with a loading phase of variable duration $t_\mathrm{load}$ where the 390~nm PI laser is pulsed on. The 423~nm PI laser is kept on during the full cycle. Loading is followed by a variable time $t_\mathrm{wait}$, where we either keep all settings constant or apply an rf tone at frequency $\omega_{\mathrm{tickle}}$ to electrode $e_{\mathrm{in,\;17}}$. Finally, an extraction pulse of 20~ns duration is applied to three electrodes, $e_{\mathrm{out,\;1}}$ with 14~V amplitude, and $e_{\mathrm{in,\;3,8}}$ with 10~V amplitude, which ejects trapped charges from the trap. For the experiments presented here we supply a constant current to the calcium oven, and keep the microwave trap drive in continuous-wave mode such that the voltage amplitude on the microwave resonator corresponds to about 90~V. We use about 500~$\mathrm{\mu W}$ of 423~nm laser light and approximately 2.4~$\mathrm{mW}$ of 390~nm light for the photoionization process. The mesh is at 150~V potential while the first and second stage, and the anode of the MCP are kept at 200~V, 2200~V and 2500~V, respectively.

In Fig.~\ref{fig:Fig4}(b) we show a histogram of MCP detection events, where the loading and wait times are $t_{\mathrm{load}}=t_{\mathrm{wait}}\approx~10~\mathrm{\mu s}$. Data are displayed as probability to record an event during a 1~ns time bin and we acquire data for $10^{7}$ experimental cycles. During the loading period we observe a small constant signal mirroring the shape of the 390~nm laser pulse, likely from just created but not trapped electrons. Application of the extraction pulse at the end of the experimental cycle results in a large and sharply localized signal from the MCP, demonstrating that electrons remain in the trap 10~$\mathrm{\mu s}$ after the end of the loading pulse. The inset displays a close-up of the readout signal, which peaks with a full-width at half-maximum of about 2~ns. Note the inset uses a linear scale for the readout signal, while the full cycle is displayed using a semi-logarithmic scale to show the background during loading as well.

\section{\label{results}Results}
\subsection{\label{exp_load}Electron loading and storage}
\begin{figure}
\includegraphics{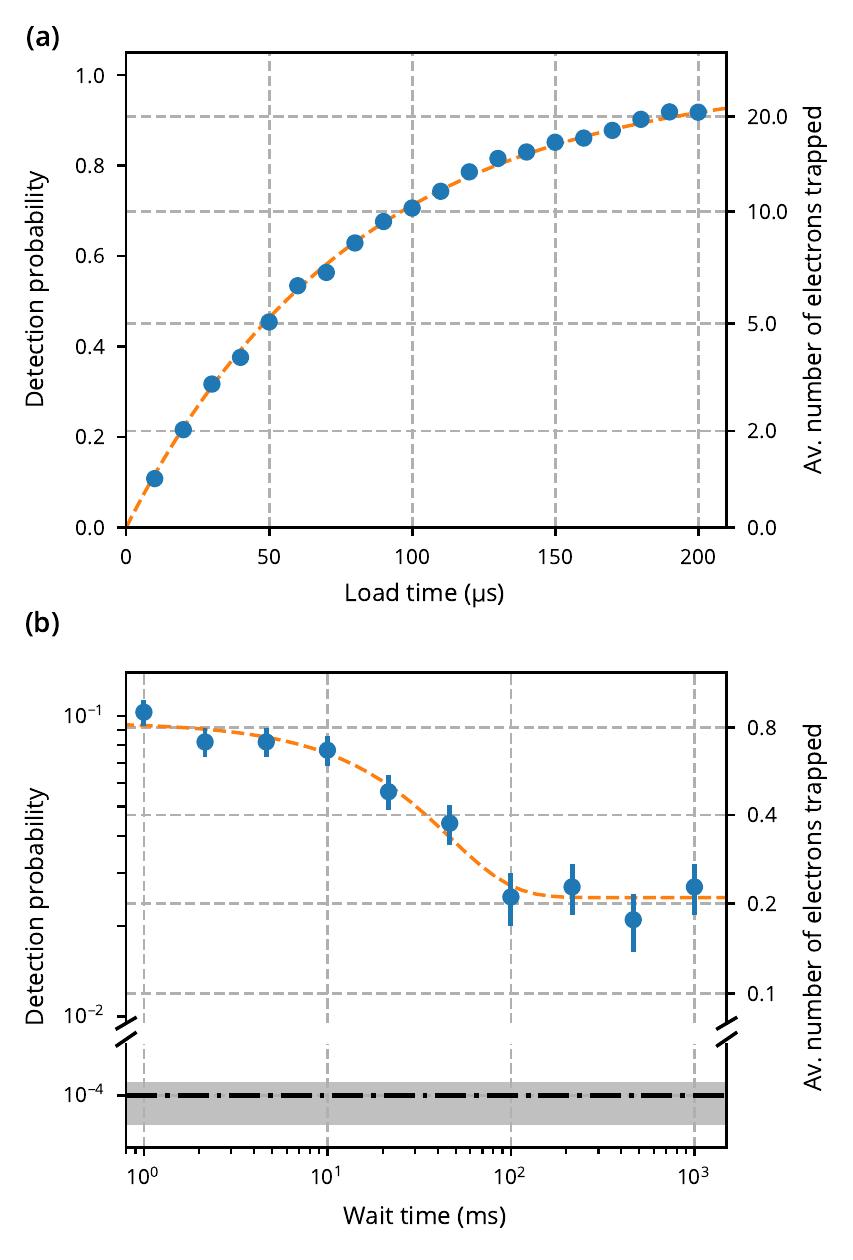}%
\caption{\label{fig:Fig5} Trapped electron loading and storage. (a) Electron trapping and detection probability as function of loading time with $t_{\mathrm{wait}} = 10~ \mathrm{\mu s}$. The dashed curve shows an exponential fit to $ 1-\exp(t/\tau_{1/e})$ with $\tau_{1/e}= 80.3\pm 0.5~\mathrm{\mu s}$. Error bars due to statistical uncertainty are too small to be visible. (b) Storage time measurement with a double-exponential fit, showing three quarters of electrons leave the trap with a decay constant $\tau_{1/e} = 30 \pm 7~\mathrm{ms}$, while the remaining quarter show no decay within measurement uncertainty. The horizontal dash-dotted line displays the background detection level based on an independent measurement. Error bars and the grey band correspond to one s. d. statistical uncertainty. }
\end{figure}

Having demonstrated electron trapping, we move on to quantify the trapping process. To investigate electron loading we use the protocol introduced in Fig.~\ref{fig:Fig4}(a) and vary $t_{\mathrm{load}}$ for a fixed wait time $t_{\mathrm{wait}} = 10~\mathrm{\mu s}$. For ease of presentation, we sum detections over a 50-ns wide window around the readout signal, see Fig.~\ref{fig:Fig5}(a) for the results. The left ordinate displays the fraction of cycles with at least one detection pulse from the MCP which approaches unity with a time constant $\tau_{1/e}\approx 80~\mathrm{\mu s}$ as $t_{\mathrm{load}}$ increases. We employ a simple threshold method to detect MCP pulses with the TDC and set a 60~ns deadtime following each detection to prevent double counting some events due to voltage ringing, which sets a a natural limit of one detection event per cycle. Considering electron loss at the mesh and the MCP we estimate about one in eight extracted electrons results in a signal from the MCP. Taking into account the loss during the readout process and the fraction of detections per cycle, we can estimate the average number of electrons in the trap for each measurement setting (see Appendix \ref{sec:app_efficiency} for details). The right ordinate in Fig.~\ref{fig:Fig5}(a) shows that the electron number is proportional to the loading time: it takes on average $10~\mathrm{\mu s}$ to load one electron and we trap on average about 20 electrons for a loading time of $200~\mathrm{\mu s}$.

To measure the electron storage time in the trap, we set the loading time such that the trap rarely contains more than a single electron, and record the readout signal as function of the wait time, see Fig.~\ref{fig:Fig5}(b) for the data. The measurement shows two distinct regimes, where about three quarters of electrons are lost within 100~ms (exponential decay with $\tau_{1/e} = 30 \pm 7~\mathrm{ms}$ for this measurement), while the remaining one quarter show no detectable loss after 1~s. The dark horizontal dash-dotted line displays the background detection level measured independently to be about $1\times 10^{-4}$ detections per cycle.

Long storage times in the trap are essential if trapped electrons are to be used as qubits, so understanding the mechanism behind loss is an important task.
The dominant loss mechanism for laser-cooled ions is collisions with background gas. Collisions, in particular with heavier atoms and molecules, can provide sufficient energy to kick an ion out of its trapping potential or may lead to the formation of molecules. Given the light mass of the electron one might expect collisional loss and electron capture to play a major role in our trap too.
In order to quantify this loss channel we have changed the pressure in the vacuum chamber by more than an order of magnitude and measured the decay constant at chamber pressures of about $\sim5\times10^{-10}$~mbar and $\sim2\times10^{-8}$~mbar. We found no change within our measurement uncertainty (see Appendix \ref{sec:app_loss} for details), which rules out collision with the background gas as the primary loss channel.

The important difference between electrons in our trap and laser-cooled ions is their light mass as well as the absence of a cooling mechanism which dampens the motion and concentrates particles in the trap center. Electrons in our trap sample a much greater volume and hence experience anharmonicities in the trapping potential further away from the trap center. While the general motion of a charged particle in an anharmonic AC potential is non-trivial~\cite{Gerlich1992}, we can make two general distinctions to the harmonic case. First, the normal modes of motion become coupled and second, nonlinear resonances \cite{Wang1993, Alheit1996a} enable the transfer of energy from the driven micromotion to the secular motion, leading to heating and particle loss. We illustrate particle loss for the motion of a single electron along one dimension of the trap in numerical simulations in Appendix \ref{sec:simulation} and find that trajectories for electrons ionized further than about 100~$\mathrm{\mu m}$ from the trap center are not generally stable. The electron motion amplitude in the trap exceeds 250~$\mathrm{\mu m}$ for those cases. This length scale is consistent with the onset of strong deviations of the pseudopotential from a purely harmonic form as shown in Fig. \ref{fig:Fig2}. Within $200~\mathrm{\mu m}$ of the trap center, the pseudopotential deviates by less than 2\% from the ideal harmonic potential and one may expect stable motion there. Coupling between the secular modes and repeated perturbations can still heat electrons from stable into unstable orbits and lead to loss, however, since there is no damping of the motion. Such perturbations to the trapping potentials do happen in our experiments, for instance due to power fluctuations of the microwave field, charging of the trap substrate, or the interactions of multiple charges in the trap.

We can experimentally probe some elements of this argument: by changing the focusing and alignment of the PI beams, which affects the average ionization distance from the trap center, we find larger PI beams in the trapping region are correlated with fewer long-lived electrons. We also observe storage times to decrease slightly when we increase the electron density in the trap by increasing the loading time. 

We believe trap anharmonicity then to be the driving force behind electron loss in our trap. Only the trajectories for single electrons confined within the harmonic trapping region are stable, so with the ionisation volume dictated by the alignment of the photoionization beams we expect the majority of initially trapped electrons (which are created outside of the central $100~\mathrm{\mu m}$) to experience a sufficiently anharmonic potential be driven out of the trap eventually. The tail of long-lived electrons in Fig.~\ref{fig:Fig5}(b) is then attributed to the single electrons that remain in a stable trajectory after all other particles have been heated out of the trap.

Studying the loss mechanisms of trapped electrons in greater detail with this Paul trap will likely be an important subject for future work. However, the long lifetimes observed here show that heating effects, for instance due to collisions with background gas or the electron micromotion~\cite{Prestage1991, Chen2013a} are not prohibitive to conducting experiments even at room temperature. This is an encouraging sign, in particular for the prospects of non-destructive electron detection and cooling via image current measurements in a cryogenic environment.
Studying trap loss may also yield insights relevant for quadrupole ion traps, and since the electron motion is faster by a factor of a hundred compared to ion motion in a typical trap, experiments would take less time.

\begin{figure}
\includegraphics{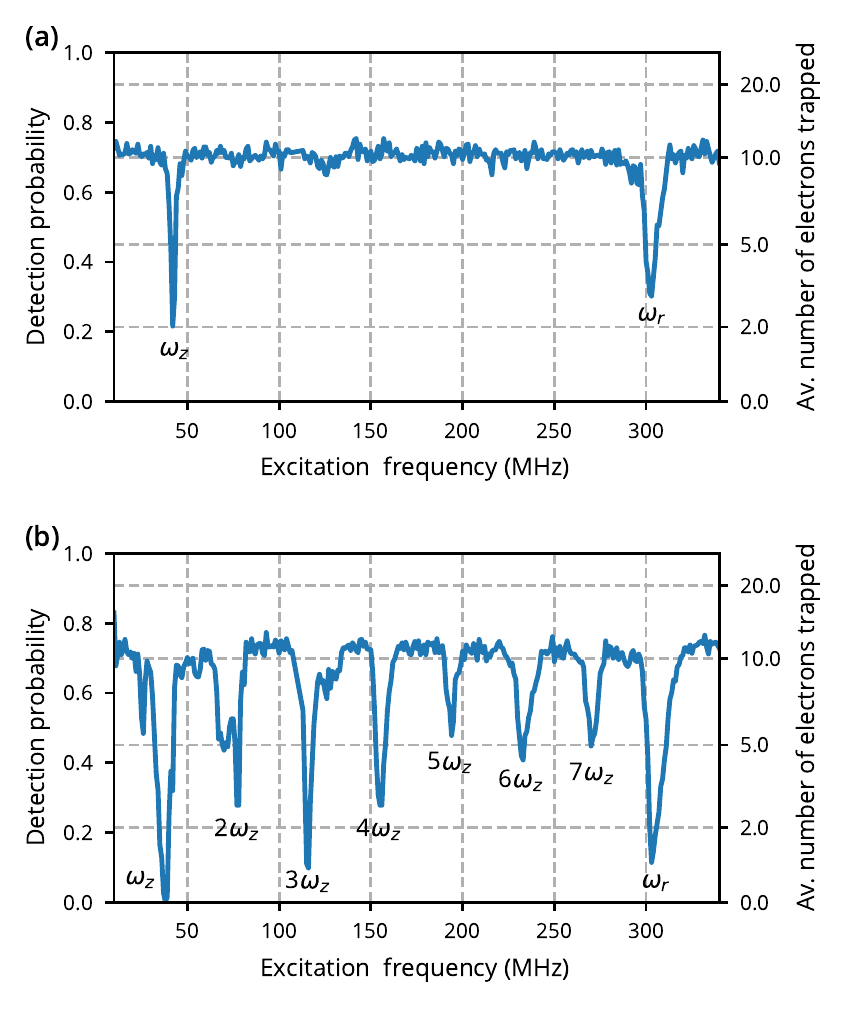}%
\caption{\label{fig:Fig6} Trap frequencies. Measurements of the motional resonances for an average of four electrons loaded into the trap, and a wait time $t_{\mathrm{wait}} = 2~ \mathrm{ms}$ during which an rf tickle is switched on. The axial (radial) resonance is denoted $\omega_{z}$ ($\omega_{r}$). (a) 5~mV tickle excitation, (b) 20~mV tickle excitation.}
\end{figure}

\subsection{\label{frequencies}Trap frequencies}
Finally, we are interested in the frequencies of the electron motion in the trap. Again, we follow the experimental protocol from Fig.~\ref{fig:Fig4}(a), now loading about ten electrons on average, and setting the wait time to $t_{\mathrm{wait}} = 2~\mathrm{ms}$. During the wait time we try to excite the motion of trapped electrons with an rf tone at frequency $\omega_{\mathrm{tickle}}$. We step $\omega_{\mathrm{tickle}}$ in increments of 1~MHz from 20 to 350~MHz and monitor electron loss, which is indicative of a motional resonance. The top panel in Fig.~\ref{fig:Fig6} shows the electron loss spectrum for a tickle voltage of about 5~mV applied to $e_{\mathrm{in,\;17}}$ and features two prominent dips. We can identify the resonances based on their response to DC and microwave voltages, revealing the dip at about $2\pi\times 40$~MHz as the axial mode, while the $2\pi\times 300$~MHz resonance is due to one of the radial modes of motion. We find no evidence of multi-electron Wigner crystal modes, which is consistent with having weakly or non-interacting electrons with a range of energies in the trap.

For low excitation tickle powers only the fundamental resonances are visible. Exciting the system more strongly reveals a series of harmonics of the axial mode (see Fig.~\ref{fig:Fig6}(b)), and a small shift in the fundamental frequency.
Changing the DC and microwave voltages, we can tune the axial mode frequency between 30 and 100~MHz and the radial mode between 200 and 380~MHz, limited by the voltage sources used in the experiment.

\section{Conclusions and outlook}
In summary, we have presented the first experiment to trap electrons in a microwave Paul trap. Electrons can be loaded in tens of microseconds and 25\% survive up to at least one second. Trap frequencies ranging from several 10~MHz to several 100~MHz have been measured.

Trapping electrons in a Paul trap opens to door to using their unique properties for quantum information processing.
One of the main challenges moving towards this goal is to cool the secular modes sufficiently to be able to perform quantum operations on them. The well-established method of detecting the electron image current with a resonant circuit provides a convenient cooling mechanism, as it thermalises the detected mode of motion with the temperature of the detection circuit~\cite{Wineland1973}. Some form of cryogenic cooling is then necessary to reach low electron temperatures. This may not appear straightforward considering the high voltages at microwave frequencies required for trapping, but several features may be used to our advantage. First, we do not, in principle, require the trap itself to be cold, just the detection circuit, which can be separate from the main body of the trap. Since the image current detection dissipates much less power than the microwave electrodes, cooling the detection circuit separately may be a sensible choice. Second, we do not not need to reach the motional ground state to be in the quantum regime~\cite{Molmer1999}.
For the sake of concreteness, reducing the dimensions of the current trap by a factor of ten, such that the trap center would be about $50~\mu\mathrm{m}$ from the nearest electrode, and increasing the drive frequency by a factor 10 ($\Omega\sim16~\mathrm{GHz}$), would give radial modes at $\omega\sim3~\mathrm{GHz}$. With a detection circuit cooled to 1.5~K and tuned to a radial mode, we expect an average mode occupation of 10 quanta, which is comparable to the mode occupation of a Doppler-cooled ion. In Ref.~\cite{Peng2017} we have discussed schemes to cool other modes of motion and perform state readout under similar conditions.

Along these lines, the next milestones towards quantum control of trapped electrons would be non-destructive electron detection~\cite{Wineland1973} and spin readout~\cite{Peng2017}, which benefit from motional frequencies in GHz regime (that is, smaller traps) and integration into a cryogenic environment. Building on technology that has already been demonstrated for quantum control of trapped ion hyperfine~\cite{Mintert2001a, Piltz2016, Ospelkaus2011, Harty2016} and Zeeman~\cite{Ruster2016} qubits could accelerate the development of a trapped electron quantum computing platform.
Distribution of entanglement over large distances is another challenge further in the future. Here, dipole-dipole coupling of single electrons, or electron crystals, in separate traps is an attractive option to realizing entanglement over intermediate distances~\cite{Harlander2011,Brown2011}. A path towards coupling electron qubits over longer distances could be via image currents in shared electrodes~\cite{Heinzen1990,Daniilidis2009}. Techniques like these may enable creation of large entangled states. 

We also note that an electron in the harmonic potential of a Paul trap realizes an instance of the lightest possible electromechanical oscillator~\cite{Eisert2004}. The resonance frequency and quality factor can be engineered by controlling the confining potential. While we believe a platform operating with trapped electrons as the sole qubit modality is the least challenging route, the ability to fine-tune the frequency of motion {\em in-situ} and the electron's strong interaction with electric fields could be used for coupling to other quantum systems with resonances in the GHz range, such as superconducting qubits~\cite{Kurizki2015,Kotler2017}.
Finally, electron Paul traps may also find applications outside the realm of quantum information science. Our trap could, for instance, trap positrons and be employed for the preparation of antihydrogen~\cite{Leefer2016,Ahmadi2020}.
Other applications include electric-field sensing at GHz frequencies~\cite{Brownnutt2015}, using cold trapped electrons for imaging~\cite{Kruit2016}, or for plasma physics studies~\cite{Twedt2012}.

\begin{acknowledgments}
We would like to thank S. Mouradian for manuscript feedback, the Yao lab at UC Berkeley for loan of the TDC, and Dr David E. Root, Keysight Laboratories, for initiating financial support and technical assistance by Keysight Technologies through the Keysight University Research Collaborations Program. 
\end{acknowledgments}

\appendix
\section{\label{sec:app_efficiency} MCP detection efficiency}
\begin{figure}
\includegraphics{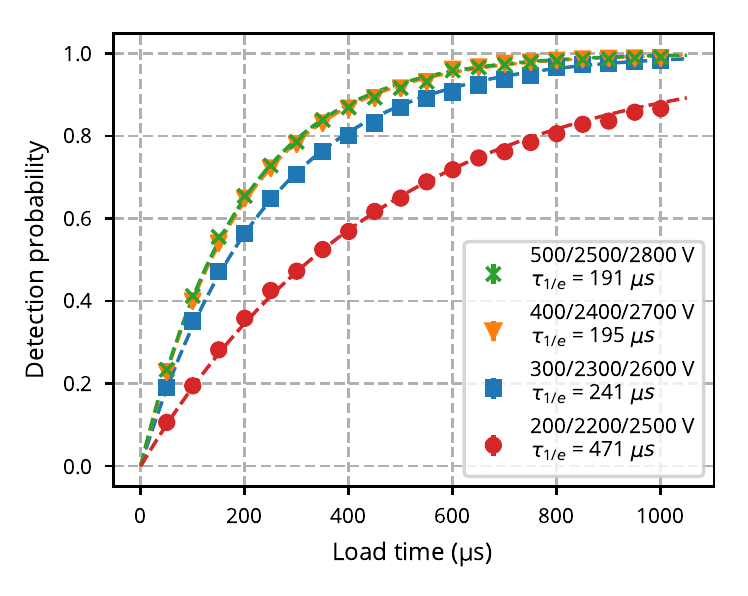}%
\caption{\label{fig:FigA1} Calibration of relative MCP detection efficiency. Measurement of electron loading rates (compare to Fig. \ref{fig:Fig5}(a)) for a range of MCP voltages. For typical operating conditions in this work (200/2200/2500~V) the detection efficiency is about 40\% of the maximum efficiency.}
\end{figure}
To be registered in our readout scheme electrons have to escape the trap, pass the mesh, impinge on the MCP, and trigger the electron multiplication process. From electron trajectory simulations we estimate the extraction process to be close to lossless, and hence the open area of the mesh and the non-unity MCP efficiency are the main loss factors. The open area of the mesh is 0.5. The MCP detection efficiency depends on the energy of incoming electrons, which is determined by the potential difference between trap and the first MCP stage, and the voltages we apply to the second stage and the anode of the MCP. For sufficiently high MCP voltages and electron energies of a few hundred eV, the efficiency should saturate approximately at the MCP open area of about 0.6, such that the upper bound on electron detection efficiency would be about $1/3$. We performed electron loading measurements like the one shown in Fig. \ref{fig:Fig5}(a) with a range of MCP voltages to calibrate the MCP detection efficiency for the operating conditions used in the main body of the paper.

Figure \ref{fig:FigA1} shows the data (symbols) and fits to $1-\exp(t/\tau_{1/e})$ (dashed curves). The legend displays the MCP voltages and the fit parameter $\tau_{1/e}$. We note that the $\tau_{1/e}$ values are different from the Fig. \ref{fig:Fig5}(a) data as the intensity of the photoionization beams was lower here. We observe a clear saturation of the detection efficiency as the voltages approach 500~V, 2500~V, 2800~V for the first stage, second state, and anode of the MCP, respectively. For these voltages the MCP efficiency should correspond to the open area of 0.6 and it follows that for the typical operating conditions used here the efficiency is reduced by a factor of about 2.5. Overall, we then estimate that 12\% of electrons leaving the trap in the extraction process generate a detectable signal from the MCP.

A caveat in this estimate is that we are not accounting for possible degradation of the MCP which would lower the efficiency somewhat.

\section{\label{sec:app_number} Electron number estimate}
The conversion of measured detection probability to electron number assumes that electron detection events at the MCP follow Poissonian statistics. Concretely, we find the mean $\lambda$ for a Poisson distribution $P_{\mathrm{P}}(X, \lambda)$ such that $P_{\mathrm{P}}(0, \lambda)$ matches the measured probability to record no events during readout $P(!\mathrm{detection}) = 1-P(\mathrm{detection})$. The product of $\lambda$ and the electron loss (inverse of detection efficiency) provides an estimate of the electron number in the trap at the point when the extraction sequence is applied.

\section{\label{sec:app_loss} Supplemental details on electron loss}
\subsection{Influence of collisions}
\begin{figure}
\includegraphics{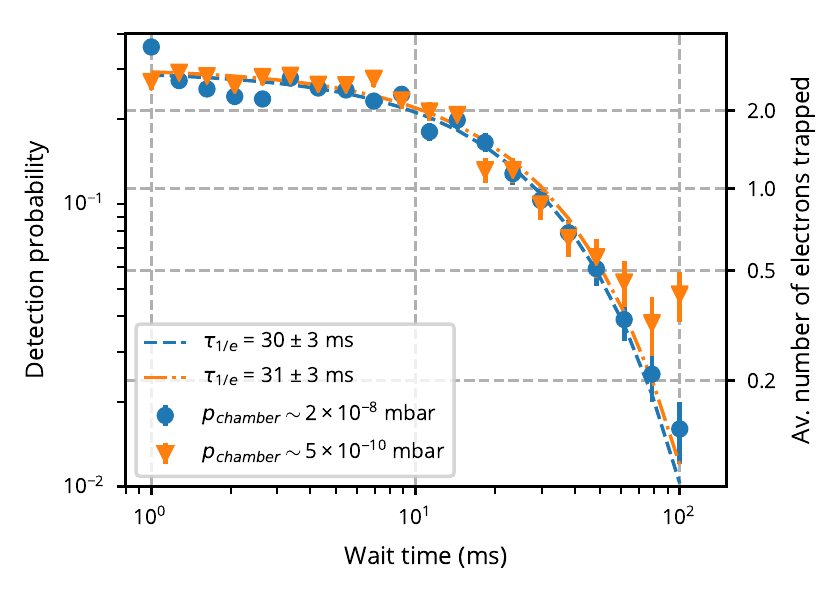}%
\caption{\label{fig:FigC1} Electron storage measurements with different background pressure. Data shown as filled circles (fitted by dashed curve) are taken at typical operating conditions, where the ion pump is switched off. Data shown as triangles (fitted by dash-dotted curve) are taken at lower chamber pressure, which is achieved by running the ion pump.}
\end{figure}
An ion pump with line of sight to the main vacuum chamber is used to keep the pressure below the $1\times10^{-10}$~mbar level. We found that operating the MCP while the ion pump is switched on leads to background detections in excess of 10~kilocounts per second. When the Ca oven is hot and the trap microwave drive is on, the pressure in the chamber increases to $\sim5\times10^{-10}$~mbar and the electron detections due to the ion pump increase by 1-2 orders of magnitude, lowering the signal to noise ratio of our measurements. Hence, we typically operate with the ion pump switched off during measurements. The pressure in the chamber then increases above $1\times10^{-8}$~mbar.

Fig. \ref{fig:FigC1} shows two electron storage time measurements. The data shown as circles are taken under typical conditions, where the ion pump is switched off and the chamber pressure is $\sim 2\times10^{-8}$~mbar. Turning on the ion pump and repeating the measurement we record the data shown as triangles. To account for the detections due to the ion pump, the background level is measured separately and subtracted from this data set. As the storage times agree within the measurement uncertainties we rule out collisions with background gas as the dominant loss mechanism.

\subsection{\label{sec:simulation}Stability of electron motion}
\begin{figure}
\includegraphics{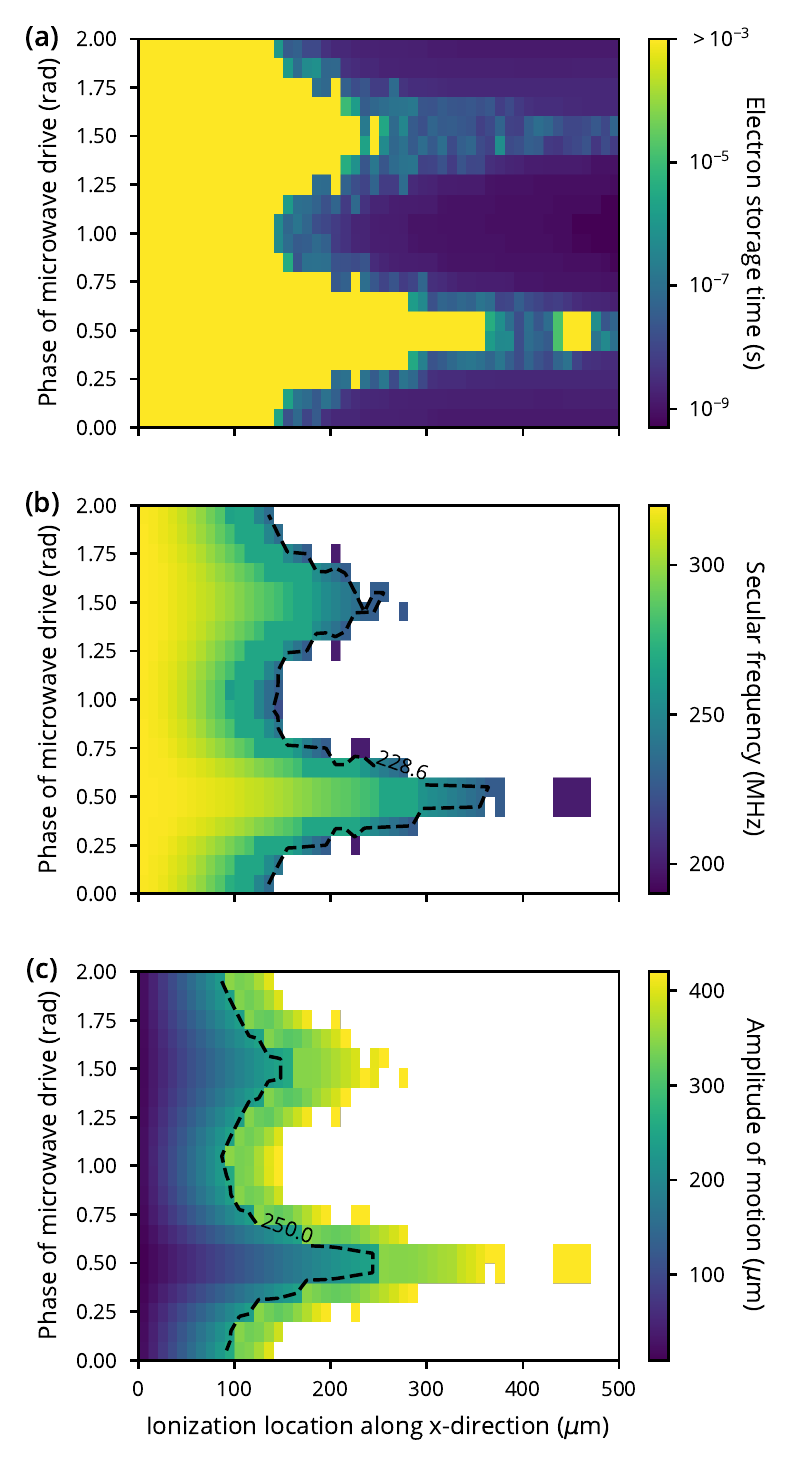}%
\caption{\label{fig:FigC2} Electron motion simulations. (a) Storage times for one-dimensional motion through the trap along the $x$-axis and the trap center (see Fig. \ref{fig:Fig2}(a), (b)). Storage times are calculated for a maximum time of 1~ms. The plot axes indicate the ionization location and the microwave phase at the time of ionisation. The electron is assumed to have zero kinetic energy at the time of ionisation. (b) Secular frequency of motion. The dashed curve highlights the frequency contour for the $7^{\mathrm{th}}$ subharmonic of the microwave drive frequency. (c) Corresponding motion amplitudes. The dashed contour line for an amplitude of 250~$\mathrm{\mu m}$ highlights where the motion frequency coincides with the $6^{\mathrm{th}}$ subharmonic of the drive frequency and the motion amplitude increases abruptly. White areas in (b) and (c) indicates storage times below $1\mathrm{\mu s}$.}
\end{figure}
Here, we evaluate the stability of electron trajectories in the trap by numerical integration of the electron motion. As discussed earlier, the energy, and consequentially the trajectory, of a single electron is determined by the surplus ionisation energy, the ionization location and the phase of the microwave field at ionization. Our primary aim will be to illustrate the electron loss mechanisms rather than attempting to fully reproduce our experiments in simulation, so we simplify the computational problem by considering motion along just one axis and ignoring the energy due to the ionization process. The latter is likely small compared to the energy of the pseudopotential and will not advance our understanding much. The path of the ionization lasers follows the $x$-axis when projected on the $xy$-plane of the pseudopotential (cf. \ref{fig:Fig2}(a)) and, based on neutral calcium fluorescence measurements we know electrons are created up to several millimeters away from the trap center. Hence, we choose to simulate the motion of a single charge in the time-dependent field from the microwave electrode along the $x$-axis. The simulation variables are the ionization distance from the trap center and the phase of the microwave drive at the ionization time.

Fig. \ref{fig:FigC2} displays the simulation results. The three panels share $x$ and $y$ axes. In Fig. \ref{fig:FigC2}(a) we show a map of the storage time in the trap. The calculated storage time is capped at 1~ms to keep calculation times reasonable. Trajectories that live up to this time are likely stable indefinitely in our simulation. We see that electron trajectories are universally stable for ionization distances up to about $120~\mathrm{\mu m}$ from the trap center. Beyond that distance the storage time depends strongly on the microwave phase, ranging from sub-nanoseconds to more than 1~ms. The transition from stable to unstable trajectories is very pronounced for all microwave phases.

\begin{figure}
\includegraphics{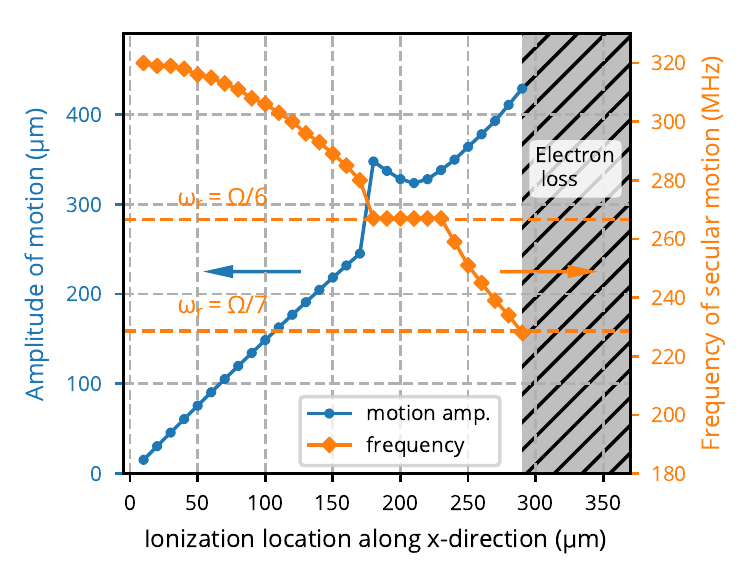}%
\caption{\label{fig:FigC3} Amplitude (left axis, blue dots) and frequency (right axis, orange diamonds) of electron motion as function of ionization location for a microwave drive phase of $\phi = 0.35\times\pi$ at ionization. The orange dashed lines indicate subharmonic resonance frequencies. Electron motion becomes unstable when the secular frequency reaches the $7^{\mathrm{th}}$ subharmonic, corresponding to a motion amplitude of about 420~$\mathrm{\mu m}$. For ionization distances from the trap center of about 180-230~$\mathrm{\mu m}$ we confirm that the subharmonically driven motion is accompanied by an increase in the motion amplitude.}
\end{figure}

The reason behind this striking behavior becomes clearer when we look at the frequency of the secular motion and its amplitude, which are presented in Fig.~\ref{fig:FigC2}(b) and~(c). Frequencies and amplitudes are only shown where the storage time in the trap exceeds $1~\mu\mathrm{s}$, white areas indicate more rapid electron loss. The transition to unstable trajectories in Fig.~\ref{fig:FigC2}(b) happens at the same secular frequency for all parameter combinations (see dashed contour line) and we can identify this frequency as the $7^{\mathrm{th}}$ subharmonic of the microwave drive frequency: $1.6~\mathrm{GHz}/7 \approx 229~\mathrm{MHz}$. The $6^{\mathrm{th}}$ subharmonic around $267~\mathrm{MHz}$ is also visible as a uni-colored band in parameter space where the secular frequency is locked to the subharmonic (see also Fig. \ref{fig:FigC3}). At these subharmonic frequencies, energy is pumped from the driven micromotion into the secular motion, which heats up the motion. We can observe the heating effect on the map of motion amplitudes in Fig.~\ref{fig:FigC2}(c): the motion amplitude first increases smoothly as function of the ionization distance from the trap center, until an amplitude of about 250~$\mathrm{\mu m}$ is reached (see labeled contour). Here, the motion amplitude suddenly increases. This jump coincides with the secular frequency hitting the $6^{\mathrm{th}}$ subharmonic. The trap anharmonicity is not sufficient to drive electrons out of the trap, however. Electron loss occurs for motion amplitudes just above 400~$\mathrm{\mu m}$, when the $6^{\mathrm{th}}$ subharmonic is reached.

The interplay of motion amplitude and frequency is more easily visualized when we pick a single phase of the microwave drive. Fig.~\ref{fig:FigC3} displays the motion amplitude (blue dots, left axis) and the secular frequency (orange diamonds, right axis) for a microwave phase at ionization of $\phi = 0.35\times\pi$. The $6^{\mathrm{th}}$ and $7^{\mathrm{th}}$ subharmonic frequencies are indicated by the dashed lines. As described before, the frequency locking to the $6^{\mathrm{th}}$ subharmonic is accompanied by a sudden increase in the motion amplitude and electron loss occurs when the secular frequency coincides with the $7^{\mathrm{th}}$ subharmonic of the drive frequency.

The analysis of one-dimensional electron motion illustrates the relevance of nonlinear resonances for anharmonic potentials. On a qualitative level it is straightforward to extrapolate from one to three dimensions. Since all modes of motion are coupled, the density of nonlinear resonances in frequency space inflates~\cite{Alheit1996a}. Energy can be pumped from the driving field into the electron secular motion at a wide range of frequencies.
While many electron trajectories are stable in simulations, like for the $6^{\mathrm{th}}$ subharmonic discussed earlier, the trajectories are not robust to perturbations when a cooling mechanism is not present at the same time. Perturbations to the motion, for instance due to space charge effects of nearby untrapped or trapped electrons, or charges on the trap itself, or fluctuations in the power of the microwave drive can eventually push an electron into a nonlinear resonance that heats it out of the trap. We believe this process to be at the origin of electron storage times in the millisecond-range in our measurements.

Finally, Fig. \ref{fig:FigC2} (and Fig.~\ref{fig:FigC3}) can also be used to link the width of the resonances observed in the trap frequency measurements to the possible range of electron motion amplitudes. The full width of the radial mode in Fig. \ref{fig:Fig6}(b) is about 30~MHz, which is roughly the range of frequencies for an electron ionized $\sim100~\mathrm{\mu m}$ from the trap center, corresponding to a motion amplitude $\sim250~\mathrm{\mu m}$.

%\bibliographystyle{unsrt}
\bibliography{references}

\end{document}